# Enterprise System Lifecycle-wide Innovation

*Research-in-Progress*

**Sachithra Lokuge**                              **Darshana Sedera**
Queensland University of Technology        Queensland University of Technology
s.lokuge@qut.edu.au                            d.sedera@qut.edu.au

## Abstract

Enterprise Systems purport to bring innovation to organizations. Yet, no past studies, neither from innovation nor from ES disciplines have merged their knowledge to understand how ES could facilitate lifecycle-wide innovation. Therefore, this study forms conceptual bridge between the two disciplines. In this research, we seek to understand how ES could facilitate innovation across its lifecycle phases. We associate classifications of innovation such as radical vs. incremental, administrative vs. technical innovation with the three phases of ES lifecycle. We introduce Continuous Restrained Innovation (CRI) as a new type of innovation specific to ES, considering restraints of technology, business processes and organization. Our empirical data collection at the implementation phase, using data from both the client and implementation partner, shows preliminary evidence of CRI. In addition, we state that both parties consider the implementation of ES as a radical innovation yet, are less interest in seeking further innovations through the system.

## Introduction

Market forces, technological advancements and the scientific frontiers have made innovation a necessity for contemporary organizations (Amabile 1996; Thomson and Webster 2013). Cross disciplinary scholars, in social psychology (Eisenberger et al. 1990), information systems (Rajagopal 2002), management (McAdam and Galloway 2005) and marketing (Belz et al. 2007) have agreed with the criticality of innovation for organizational continuous well-being (Boer and Gertsen 2003). For many organizations, innovation within is attained through dedicated research and development facilities (Davenport 1993), while some attempt outsourcing innovation through external service providers (Mahnke and Özcan 2006). Over the past several decades organizations have sought to adopt Enterprise Systems (ES) as a way to innovate organizational processes and structures (Bradford and Florin 2003). These systems purportedly include best-practices of doing business that allow organizations to radically change their existing processes and organizational structures (Legare 2002). Though the outcomes of initial stream of ES implementations were questionable in terms of on-time, on-budget completions, and almost all academic studies on ES implementation reasons explicitly or implicitly recognized *innovation* as an expected outcome (Karim et al. 2007). Implementation of an ES is a complex strategic issue that has a high impact on core business functions of the organization, thus; the initial ES implementation considered as a radical innovation (Kraemmerand et al. 2003), which is characterized through substantial changes to business processes (Bingi et al. 1999) management structures (Aladwani 2001) and expected outcomes (Karim et al. 2007).

In recent times, practitioner and academic studies have discussed the longevity of ES innovation (Kemp and Low 2008; McAfee and Brynjolfsson 2008). Specifically, they questioned the *readiness* of the organization, client and the vendor for continuous innovation for the Lifecycle-wide success of the ES. Considering that ES rarely replaced or retired (Eden et al. 2012), organizations must seek to innovate continuously through such applications. Research on ES use (Botta-Genoulaz and Millet 2005) and ES benefits (Chou and Chang 2008) all highlight that innovation through ES seems to diminish after the initial implementation.

Simultaneously, vendors and consultants too are under growing pressure to engage in lifecycle-wide innovations to demonstrate the delivery of benefits to the client organization (Hislop 2002). Such initiatives too impacted by the readiness of the client organization for lifecycle-wide innovation. Furthermore, market capital availability, resource constraints such as skill-shortage too have minimized the ability to gain continuous innovation through ES (Srivardhana and Pawlowski 2007). Evidencing this, Eden et al. (2012) identify that, despite the increase in the number of ES implementations, their individual budgets have shrunk in the past 10 years.

Several researchers also have observed why organizations do not receive anticipated Lifecycle wide benefits through their ES investments (Srivardhana and Pawlowski 2007). Some suggested that ES implementation Critical Success Factors (CSF) should be considered for the entire lifecycle (King and Burgess 2006), not just during the implementation. For example, CSF of implementation like (planning for) resource availability, (continuous) top management support, (continuous) knowledge management could be considered essential for lifecycle-wide innovation, where the adjective in the brackets demonstrates their applicability to the lifecycle.

Aforementioned factors highlight the need for a strategic innovation plan beyond the implementation phase. We argue that innovation should not be limited to the introduction of an ES, but something that allows continuous innovation throughout the lifecycle. Such an approach of innovation would certainly benefit the client, where the flow of benefits through ES innovation would be continuous. It is also beneficial to the software vendor and the implementation partners since the opportunities of lifecycle-wide engagement are assured.

ES, often termed, as *cement* by the practitioners (Davenport 2000), are not in general, designed it with *flexibility* in mind. Gable et al. (2008) and Sedera and Gable (2010) suggested that lack of flexibility hinders the growth opportunities of ES. Similarly, organizations rarely execute lifecycle-wide plans for continuous improvements for two main reasons (Srivardhana and Pawlowski 2007). First, organizations are frustrated by the rigidity of ES that do not encourage dynamic changes. Second, the reputation of ES in under-delivering benefits discourages allocation of further organizational resources for non-essential activities.

ES is treated as a strategic asset with great potential for innovation at the outset (Kraemmerand et al. 2003), yet, left untouched until its next scheduled upgrade. The innovation sought is continuous, yet the financial and system boundaries make it restrained. Thus, we introduce this new notion of *continuous-restrained innovation*.

The three specific objectives of this paper are; first, we position implementation of ES as a concept of innovation. Though ES is considered as an enabler of innovation, seldom they characterize ES through innovation classifications. Therefore, our understanding of ES as an innovation is still superficial. Second, using theoretical prepositions of past innovation studies; we demonstrate the adequacy of the current innovation types to explain the ES lifecycle as per Markus and Tanis (2000). Third, we propose Continuous-Restrained Innovation (CRI) as a lifecycle-wide innovation concept and we sought the level of awareness of client and implementation partners on this new phenomenon. In this study, we consider ES as an enabler of lifecycle-wide innovation and through ES, organizations can attain increased productivity and efficiency in their business processes.

We commence our paper by introducing innovation, its core attributes and we posit ES as an innovation. Third, we explain the motivation for CRI. Then, we demonstrate how the phases of the ES lifecycle can be viewed through different types of innovation. Finally, we discuss the preliminary study results followed by the conclusion, limitations and the future work.

## ES as an Innovation

Researchers have conceptualized innovation in many ways (Knight 1967; Tilton 1971). Some scholars consider innovation as a discrete product or an outcome (Meyer and Goes 1988) and some contemplate this as a process (Knight 1967). Innovation can be defined through multiple levels, such as, individual, project, organizational, industry level and national level (Frambach and Schillewaert 2002). In addition, researchers have attempted to describe innovation by understanding the antecedents (Individual, organizational factors), attributes of innovation (Relative advantage, complexity, compatibility, etc.) stages of innovation (Initiation, adoption) and, typologies (product, process, technological etc.). Yet, innovation is considered as a complex subject due to its mystique nature of creation and adoption within an organization (Van de Ven 1986).

According to Zaltman et al. (1973) innovation can be defined as; "any idea, practice, or material artifact perceived to be new by the relevant unit of adoption" (p. 29). As explained by Lai et al. (2009) innovation needs not to be a totally new concept to the world. It is also can be an imitation of something already used elsewhere, but new to the unit of adoption.

In studying innovation through ES (or on the broader topic of innovation through Information Technology), past studies have employed traditional innovation concepts (Rajagopal 2002; Siau and Messersmith 2003). Though, they added a wealth of cumulative knowledge to the discipline, most innovation studies in Information Systems (IS) have assumed that resources, both human and financial, are adequate and innovation is delivered through standard specifications of delivery (i.e. contract of delivery) (Bradford and Florin 2003). In the real business world, no resource is adequate. Weeks and Feeny (2008) pointed out that the clients expects the vendors as the strategic partners for achieving organizational success. They expect these strategic partners to go beyond the specifications and innovate for the survival of the dynamic business world. In addition, Weeks and Feeny (2008) stated that clients expect three categories of innovation through information technology; (i) IT Operational innovation (e.g. email platforms, hardware) (ii) Business process innovation (e.g. ES) and (iii) Strategic innovation (e.g. new markets). According to Davenport (1998), "embrace of ES may in fact be the most important development in the corporate use of information technology in 1990s" (p. 122). Unlike the legacy systems, ES captured reusable best practices and required the organization to undergo business process re-engineering (Wu et al. 2005). The introduction of the ES revolutionized the existing practices and introduced new behaviors to the organizational subsystems and its members (Karim et al. 2007). Thus, this risky, complicated and high resource consuming process is considered as a radical innovation in the innovation literature (Sorescu et al. 2003). Damanpour (1988) stated the radical innovation causes deeper changes in an organization, such as changes of the organization structure, roles and responsibility and simply it drastically changes the way the organization carry out the business practices. Similarly, an introduction or implantation of an ES does numerous changes in the organization. Thus, we posit introduction of ES to an organization as a radical innovation. Melnyk et al. (2013) argued that ES can be considered as an administrative innovation since it changes the organizational structure and the processes. However, in this paper we position ES as an enabler of administrative innovation. The adoption of ES promises operational, managerial, strategic, IT infrastructure and organizational wide benefits (Shang and Seddon 2000). Yet, in most organizations after implementing this radical innovation, seldom improvements of the implemented system occur. There are two major reasons limiting the ES software towards innovation, (i) ES deployments are too costly and time consuming and (ii) organizations lack the right expertise needed for innovation driven ES. Innovation is an iterative process; similarly, the implementation of the ES is not the ultimate outcome, to gain the real benefits out of ES an organization should reintroduce improved innovations throughout the ES lifecycle. However, the rigid nature of the ES and the perception of the users have made it a one off innovation until the next planned upgrade. Advent of new technologies, pressure from the competitors and for the survival of the organization, it is important to keep up with the change. Even though, the continuous innovation is required, it is difficult since the innovation needs to be achieved within the already implemented system boundaries.

Thus, we identify this continuous restrictive nature of innovation as Continuous Restrained Innovation and we define it as *the process whereby an innovation adoption unit transforms ideas, skills, resources, and technology into new/improved products, processes, or service, through effective management controls under limited conditions such as resources, cost, time, scope and congruence to achieve organizational goals.*

## Motivations for Continuous-Restrained Innovation

All the past studies have speculated, innovation conceived in a free and a resourceful environment. Yet, in ES projects, especially post-go-live, managed budget, skilled staff and tightly controlled modifications to the ES software highlight a tightly controlled phenomenon. Therefore, any innovation that takes place at the ES lifecycle must take into account the restrictive nature of innovation. We identify several motivations for continuous-restrained innovation throughout the ES lifecycle.

*Implementation:* Weeks and Feeny (2008) state that there is a strong demand by the client organizations for the consultants to innovate beyond the standard agreed specifications of delivery of a contract. In addition, they described that such innovations are difficult given that they will increase coordination complexities, lacks of alignment of risks and lacks incentives for innovations. This client expectation of innovation beyond what is agreed is fueled by growth of the small to medium consulting companies that compete with the traditional pillars, offering much better value for client organizations (Tsai et al. 2009).

*Shakedown:* At the shakedown phase, the organizational innovation is highly restricted by the boundaries of the system as well as the organizational business processes implemented through ES. As such, major or dramatic changes to the system or business processes are highly discouraged (Markus et al. 2000). Practitioner and research reports suggest (Srivardhana and Pawlowski 2007) that most organizations are disengaged with business process innovation due to the inflexibility of ES. Thus, our motivation for suggesting Continuous Restrained Innovation is appropriate for this phase as well. The salient objective of the Continuous Restrained Innovation in this phase is to facilitate much needed structure to better adoption of the new technology and its business process improvements. Though ES studies have employed much-discussed Adoption-and-Diffusion Studies (Rajagopal 2002), those studies have only focused on the initial implementation rather than the continuous adoption at the shakedown phase.

*Onwards/Upwards:* As mentioned, this phase marks an entry to a stable ES environment. As such, the turbulences of the workplace are settled and users themselves are at a stage of gaining high level of confidence and expertise. This, within the boundaries of the system, provides an ideal environment for innovation. Yet, no prior study in either in innovation or in ES domains has made specific recommendations on how to innovate and what the appropriate antecedents are necessary for innovation in this phase.

## Innovation in the ES Lifecycle

The objective of this section is to demonstrate the (i) key phases of the ES lifecycle, (ii) how innovation can be attained within each of the lifecycle phases and (iii) suggest the conceptual model for Continuous Restrained Innovation. We argue that having identified the corresponding innovation with the lifecycle phase will ultimately lead to a better understanding of how to innovate in every phase of the ES lifecycle. One important factor to note is that the degree of innovativeness varies throughout the lifecycle and at each different stage different types of innovation is possible.

Markus and Tanis (2000) suggested key phases of ES lifecycle: (i) implementation, (ii) shakedown and (iii) onwards/upwards. As Ross and Vitale (2000) suggested ES performance (individual or organization) undergoes a performance dip after the go-live and after each major upgrade (see Figure 1).

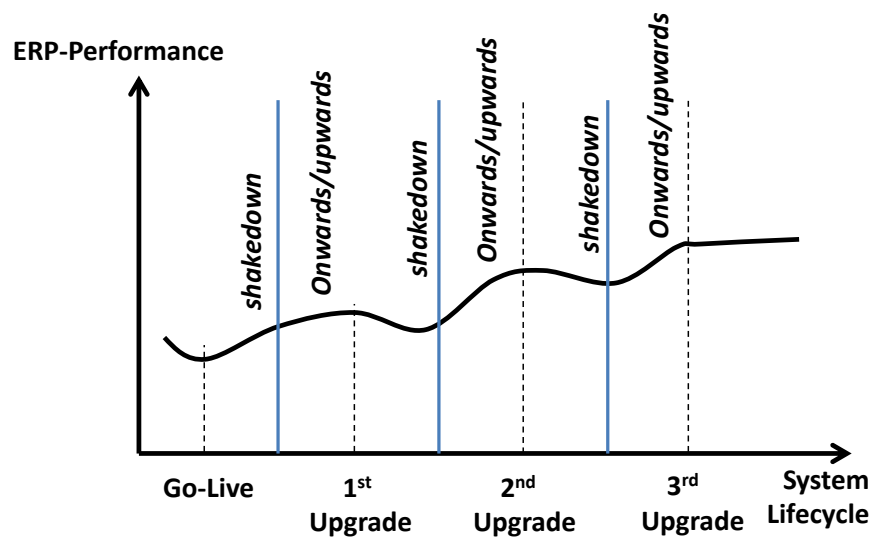

**Figure 1. Variation of the system performance in the ES Lifecycle**

Past literature have suggested that the introduction of an ES leads to innovation in the organization (Rajagopal 2002), yet empirical evidence of this have been few. An ES implementation is a consolidated effort by the client, implementation partner and the vendor, usually results in substantial changes to their existing technology, people, organizational culture and business processes of the organization (Finney and Corbett 2007). In general, the implementation partner leads the implementation process under the high-level directions of the software vendor. The implementation approach could be of either big-bang, phased or a hybrid approach (Brown and Vessey 1999).

As we argued earlier, the implementation of ES can be viewed as a radical innovation. Some studies questioning the value of ES argued that ES do not provide a strategic value or innovation, given that packaged software is available to all organizations. Yet, the definition of innovation we adhere suggests considering the unit of adoption it can be considered as an innovation.

The radicalness of the innovation are determined by the magnitude and its effect of the innovation on the organization (Gatignon et al. 2002). Therefore, the characteristics of radical innovation include the degree to which it disrupts the existing organizational practices, the high risk associated with implementing it, and the new skills and abilities it requires to process throughout the process (McDermott and O'Connor 2002). But the most enticing characteristics of radical innovation is the successful launch brings definite rewards to the organization (Leifer et al. 2001). Green et al. (1995) introduced a reliable multi-dimensional measure of radical innovation. It includes technological uncertainty, technical inexperience, business inexperience and technology cost as the four dimensions measuring the extent of radicalness. When considering these four dimensions from the clients' perspective it is clear that introduction of the ES to the organization is a radical innovation. One characteristic of radical innovation is that to get the support for radical initiatives is very difficult since it involves major changes in the organizational culture and has an immense pressure on the sub-systems and its members. From the ES literature, it is evident that for the success of implementing an ES top management support is critical. In addition, innovation literature highlights the criticality of leadership roles, team composition and the role of informal networks for a successful completion of a radical innovation project (McDermott and O'Connor 2002). Equally, in ES literature scholars highlighted the CSFs of the ES implementation project (Finney and Corbett 2007). It illustrated the need for a balanced team that consists of the best and the brightest staff, a project champion, empowered decision makers, and effective communication (Nah et al. 2001).

The shakedown phase is the period immediately after ES 'go-live' and the period after each major upgrade. During this period, usually marked as a period of chaos, ES users learn about the new system features and functions and adjust their work practices. As mentioned earlier at this phase, organizations undergo a "productivity dip" (p. 237), while gaining other productivity related improvements (Ross and Vitale 2000). Challenges at this phase could be introduced through unfamiliarity of system features and functions, new changes introduced through ES to job roles and conditions, changes to work practices and culture, software related issues and users lacking confidence to adopt new technologies over legacy systems (Nah et al. 2001; Niu et al. 2011). Herein, we argue that Administrative Innovation, as per Teece (1980) to correspond with the turbulence of the shakedown phase. The administrative innovation could facilitate a smoothening of adoption of the new technology innovations (Camisón and Villar-López 2014). According to Teece (1980)improving the administrative techniques in an organization is as important as improving the technological improvements. The active engagement with the ES during this phase will enable the users to exploit possible innovations using the system.

The onwards/upwards phase follows the shakedown phase and denotes a stable ES environment. Here the organization becomes more internally consistent and where organizations strategic orientations are consistent with internal and external environmental demands. In this phase we believe the organization has gained the business process absorptive capacity (ability of an organization to identify the value of knowledge, assimilate it, and apply it (Srivardhana and Pawlowski 2007)). Therefore, the users are able to identify the areas where improvements needed. Even though, it is difficult to make a radical innovation, there is always the freedom to reintroduce innovation. Thus, in this phase the organization has the possibility to initiate product and process incremental innovations. For example, organizations can introduce new modules to their ES, add new components such as supplier and customer management software, and can improve the system functionalities by adding plug-ins. All these introductions can be considered as innovations which changes the business processes positively. In addition, to increase the performance of the organization the management can initiate administrative innovation as well. As Weeks and Feeny (2008) state there's a possibility of introducing IT operational, business process and strategic innovations in this phase.

We believe, in the long run, organizations that seek different types of innovations are able to syndicate these innovations in new ways and will gain competitive advantage.

## Preliminary Results

We designed the study to understand elements of Continuous Restrained Innovation. For this, we selected an organization at the start of an ES implementation, where we have accessed the client and the implementation partner. The study design is of three phases, matching the three phases of the ES lifecycle. This paper reports the preliminary study findings of the *implementation phase*.

Our preliminary data was gathered from 82 members of the ES implementation team. It included 40 members from the implementation partner and 42 from the client organization. The organization (henceforth referred to as SCM-company to protect anonymity covered by the university ethics agreement) decided to implement SAP Financials and Controlling (SAP-FI/CO), Materials Management (SAP-MM), Human Capital Management (SAP-HCM) and Supply Chain Management (SAP-SCM) modules in late 2013. The objectives of adoption, time phase and the scope of the implementation is consistent with recent market surveys (Kimberling 2013). In general, similar to most organizations, SCM-company thought of ES as a long-term strategic investment. Table 1 reports descriptive statistics of SCM-company. Most respondents from the client organization represent the management or senior management level.

|  | Details |  | Details |
|---:|:---:|---:|:---:|
| **Revenue in 2012-2013** | *USD $52.6 Million* | **New modules considered** | *Sales and Distribution (2015)* |
| **Industry Sector** | *Manufacturing* | **Implementation Approach** | *Phased* |
| **Expected number of users** | *71 (Phase 1) 120 (Phase 2)* | **Level of customization** | *Medium* |
| **Expected completion (months)** | *7 months (Phase 1)* | **Size of implementation team** | *91 (50 client; 41consultant)* |

**Table 1. Company Details**

Access to management staff is essential to understand; (i) whether they see ES as a tool for radical innovation, (ii) whether ES lead to continuous innovation, and (iii) whether the client is interested in long-term continuous innovation.

Access to implementation partner is vital to understand whether; (iv) the client organization requires innovation beyond what has been agreed through contracts, (v) whether they are ready to engage with the client organization as a lifecycle-wide innovation partner, (vi) to assess whether the client organization is ready for lifecycle-wide innovation and finally, (vii) whether the client and the implementation partner recognize the restraints of innovation beyond implementation

The data collection in phase one was completed in 2014 using a survey instrument. The full implementation team was sent the survey instrument (N = 82). The survey received 80 valid responses, with 42 responses from the client and the remainder (40) from the implementation partner. The data analysis is summarized using the questions, where each question was measured using multiple items. The survey instrument for each group was modified slights for the respondent cohorts, leaving the core meaning stable.

1. Does the client / implementation partner think of the ES as a tool of radical innovation?
2. Does the client / implementation partner expect ES to lead innovation throughout the lifecycle?
3. Is the client organization interested in lifecycle-wide innovation through ES?
4. Does the client seek innovation from the implementation partner at the implementation phase, beyond the agreed contract?
5. Is the implementation partner ready to engage as a lifecycle partner to engage in creating innovation using ES?
6. Is the client ready for lifecycle wide innovation using ES as a tool?
7. Does the client / implementation partner recognize restraints of innovation?

Table 2 and Figure 2 demonstrate the descriptive summary statistics as well as determined whether the client and the implementation partner have statistically significant differences across the seven questions.

The mean scores for each question was derived using the simple averages for questions pertaining to each question, while the differences were observed using independent sample t-test.

| | | Client | IP | Sig* |
|---|---|---|---|---|
| 1 | ES implementation is a radical innovation | 5.8 | 6.1 | No |
| 2 | Expect ES to lead innovation throughout the LC | 6.3 | 4.2 | Yes |
| 3 | The client is interested in continuous LC-wide innovation | 5.6 | 6.1 | No |
| 4 | Innovation at implementation should be beyond the contract | 6.1 | 6.4 | No |
| 5 | IP readiness to engage as a lifecycle partner of innovation | 3.4 | 3.3 | No |
| 6 | Client readiness for LC-wide innovation | 5.9 | 2.3 | Yes |
| 7 | Recognize the restraints of innovation beyond implementation | 3.8 | 3.4 | No |
| | IP = Implementation Partner; LC = Lifecycle; Sig* = Significance at 0.05 | | | |

**Table 2. ES innovation views of client and implementation partner at the implementation phase**

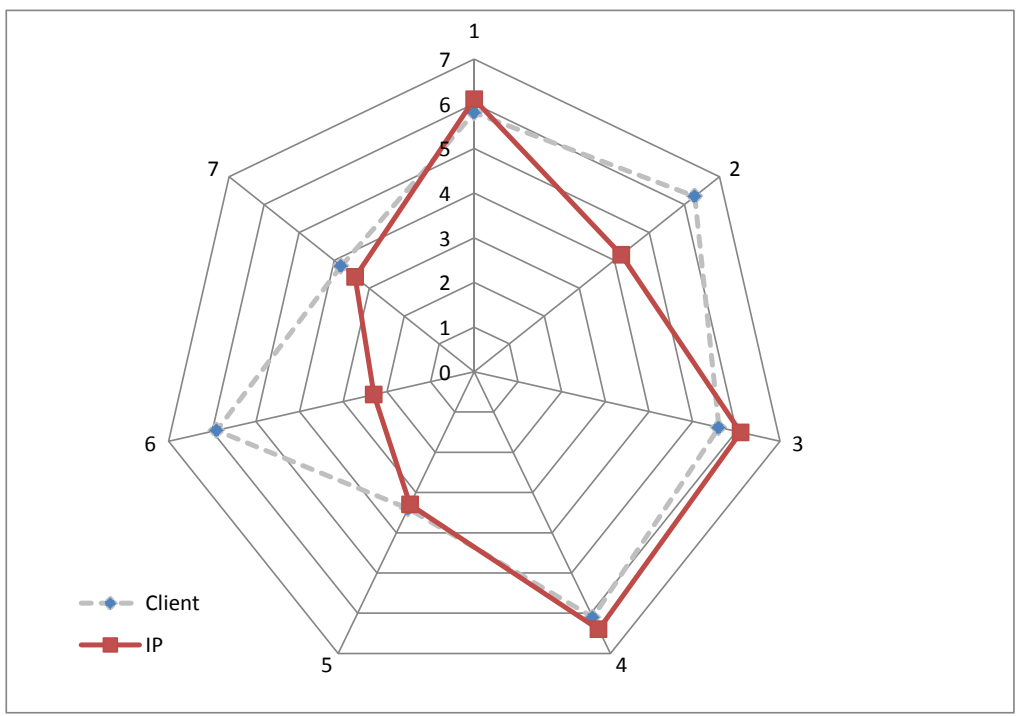

**Figure 2: ES innovation views of client and implementation partner at the implementation phase**

Using Table 2 and Figure 2, we make the following observations.

Focus 1: ES implementation is a radical innovation

With the high mean scores, it is evident that both client and the implementation partner agree that ES brings radical innovation to the organization. This is consistent with past literature of *innovation* where they argued that radical innovation depends on the unit of adoption, where the newness is observed through where it is adopted, rather than on the newness of the technology it-self. There were no significant differences between the client and the implementation partner in relation to this question.

Focus 2: Expect ES to lead innovation throughout the Lifecycle

Our preliminary results demonstrate that the client and the implementation partner disagree with the notion of lifecycle-wide innovation, with the client has a *stronger* belief that ES would lead to / facilitate lifecycle-wide innovation in the organization. This is a significant incongruence between the key partners of implementation, where such differences could lead the implementation partner to focus on delivering the ES on time and on budget without making provisions for long-term flexibility /foundation for the future.

Focus 3: The client is interested in continuous lifecycle-wide innovation

It is evident from our data that the client organization likes to engage in lifecycle wide ES innovation and the implementation partner agrees with this too. Therefore, we did not see any significant differences between the client and the implementation partner here. Yet, it was somewhat puzzling to see the differences in scores by the implementation partner for question 2 and 3.

Focus 4: Innovation at implementation should be beyond the contract

Demonstrating first evidence of Restrained Innovation, we witness that both the client and the implementation partner expect to contribute more than what is stated in the delivery contract. Here, we do not see any significant differences. This also demonstrates the competitive pressures that the implementation partner faces in contemporary ES implementations.

Focus 5: Implementation partner readiness to engage as a lifecycle partner of innovation

Here, both parties agree that the implementation partner is not ready to engage as a lifecycle-wide partner – thus we see no significant differences between their views. This result is somewhat surprising given the focus by consulting companies to push for long-term contracts, beyond ES implementations. We speculate two primary reasons for our result: (i) the client could be seeking mid-long term engagement with multiple implementation partners to mitigate risk of relying on a single consulting firm; (ii) the implementation partner does not have the capacity / resources for lifecycle-wide engagement.

Focus 6: Client readiness for lifecycle-wide innovation

We observe significant differences between the client and the implementation partner on the client readiness for lifecycle-wide innovation. Interesting observation was made using the origins of this difference between client and implementation partner views. We identified that the client organization have scored very high on the LIKERT scale for their readiness – focusing more on the availability of *financial resources*. Yet, the implementation partner, when assessing client's ability, they focus on the *knowledge resources* for lifecycle-wide engagement.

Focus 7: Recognize the restraints of innovation beyond implementation

Both client and the implementation partner recognize that innovation beyond implementation cannot be of radical in nature and that any such innovation (if at all) will be restrained by the boundaries imposed by the system, processes and organizational rules.

## Conclusion

This study sought novel perspective to innovation in ES, through adding restricted environmental perspectives. This new perspective of restrictive innovation is a reality in most IS service delivery agreements, where the client organizations expect innovation beyond the contracted terms. Despite a wealth of research on innovation from various disciplines, we were unable to find any studies researching this restrained nature of innovation throughout the ES lifecycle. Thus, this paper forms a conceptual bridge between ES research and innovation research. When visiting the ES lifecycle with innovation in mind we recognized possible innovations that can be attained throughout. However, to the best of our knowledge no scholar has exposed the restrained yet possible innovations in the ES lifecycle.

The advent of new technologies has paved many paths for organizations to perform well and at the same time, these technologies put immense pressure on the organizations to innovate for their survival.

Therefore, we argue, even though ES is a strategic initiative, organizations should focus on continuous innovation for the well-being. We believe that innovation should not be limited to the introduction of an ES, but something that permits continuous innovation throughout the lifecycle. This is a win-win for all the stakeholders of the ES. ES is considered as a major strategic investment of an organization. The implementation of and ES is considered as a very costly process, therefore, organizations seldom upgrade their ES. Yet, in the ever changing, competitive business environment, organizations cannot survive without innovating. Therefore, organizations are required to innovate within this restrictive environment. Thus, this research contributes a new concept of continuous restrained innovation to the literature.

Through our preliminary analysis, we evidenced that both client and the implementation partner consider ES as a radical innovation. Yet, there is no agreement between them on considering ES as a lifecycle-wide innovation. Our study provided initial evidence that clients consider ES as a lifecycle-wide innovation and they are ready to engage with the implementation partner for planning and managing ES as an innovation. Yet, our results evidence that the implementation partner is not ready to engage as a lifecycle-wide partner. Furthermore, our results indicate client's willingness to innovate throughout the lifecycle, yet their recognition of lifecycle-wide restraints inherited through ES. This provides first evidence of "Continuous Restrained Innovation". Given the client's choice, to survive in business, to compete and to be unique, consultants need to be more innovative. They need to focus on value addition rather than cost reduction (Weeks and Feeny 2008). Therefore, it is evident that implementation partners need to focus more on continuous innovation through the effective use of ES. Both the client and the implementation partners should not limit their relationship until the successful implementation of ES but should focus on long-term engagements. This will benefit not only the client and the implementation partners, but also the customers, vendors and all the stakeholders who will reap the benefits of the delivery of new and advanced products/services, increased productivity, profits and efficiency.

## Limitations and Future Work

Results and preliminary analysis is heartening and identifies several trends. The data is not sufficient to confirm whether these trends are new or newly discovered. Furthermore, future studies with multiple organizations are required to ascertain whether patterns observed in this study can be generalizable. Since the case organization we have taken in this study has just started their implementation process, we are planning to collect data in each stage (for example, immediately after implementation). In addition, we plan to collect data from multiple organizations and we plan to do longitudinal data analysis using these multiple organizations. We expect continuation of this project will provide us interesting insight into continuous innovation.

In this study, we have introduced few possible types of innovation that can be attained throughout the ES lifecycle. We further plan to extend our study by identifying the factors that encourage/discourage restrained innovation in the ES lifecycle. Further, we extend the study to identify the innovation readiness of clients and implementation partners and assess the innovation readiness of each involving party. In addition, we further plan to use this preliminary data to establish ES as an innovation and identify possible innovation types, considering ES as a trampoline that allows various types of innovations to stem.

We acknowledge several limitations of the study. First, a single case will hinder generalizability of the study findings. However, the current study is a representative case of a typical ES implementation in terms of the budget, number of modules and the involvement of the implementation partner. Second, self-assessment of readiness could be seen as another weakness. The third weakness relates to the method, where the survey data collection may potentially lose some richness of the study. Therefore, we extend this study by applying quantitative as well as qualitative methods.